\documentclass[amsmath, showkeys,amssymb,a4paper, onecolumn, 13pt]{revtex4}
\usepackage{latexsym}
\usepackage[dvipdfm,colorlinks=true,linkcolor=cyan,anchorcolor=cyan,citecolor=cyan]{hyperref}
\usepackage[hmargin={1.2cm,1.2cm}, vmargin={2cm,2cm}]{geometry}
\usepackage{float}
\usepackage{graphicx,amsfonts,amsmath,amssymb,amsthm,amscd}
\usepackage{epsfig}
\usepackage{bm}
\usepackage{xr}
\usepackage{indentfirst}
\usepackage{autobreak}
\usepackage{appendix}
\usepackage{changepage}

\newcommand{\dx}{\;\mathrm{d}x}

\newcommand{\difffrac}[2]{\dfrac{\mathrm{d} {}#1}{\mathrm{d} {}#2}}
\newcommand{\parfrac}[2]{\dfrac{\partial {}#1}{\partial {}#2}}
\newcommand{\Eref}[1]{Eq.\,(\ref{#1})}
\newcommand{\Fref}[1]{Fig.\,\ref{#1}}

\date{\today}

\begin{document}

\title{\bf Motility and Energetics of Randomly Flashing Ratchets}

\author{Xining Xu and Yunxin Zhang} \email[Email: ]{xyz@fudan.edu.cn}
\affiliation{Laboratory of Mathematics for Nonlinear Science, Shanghai Key Laboratory for Contemporary Applied Mathematics, Centre for Computational Systems Biology, School of Mathematical Sciences, Fudan University, Shanghai 200433, China.}

\begin{abstract}
	We consider a randomly flashing ratchet, where the potential acting can be switched to another at random. Using coupled Fokker-Planck equations, we formulate the expressions of quantities measuring dynamics and thermodynamics. Extended numerical calculations present how the potential landscapes and the transitions affect the motility and energetics. Load-dependent velocity and energetic efficiency of motor proteins, kinesin and dynein, further exemplify the randomly flashing ratchet model. We also discuss the system with two shifted sawtooth potentials.
\end{abstract}


\maketitle

\section{Introduction}
Transport of Brownian particles operating far from thermal equilibrium stays attractive during the past decades due to the wide applications \cite{LeiblerMoving, SalgerDirected, fisher1999force, gommers2006quasiperiodically, hanggi2009artificial, kawaguchi2014nonequilibrium, spiechowicz2016transient}. Typically,  a net current is produced through an combined action of asymmetry,  either spatial or temporal, and non-equilibrium arising from thermal fluctuations, leading to a directed motion \cite{julicher1997modeling, astumian1997thermodynamics, ReimannBrownian}.  It is the broken symmetry that rectifies the unbiased fluctuations, which creates the key criterion for the possibility of transport. Such mechanism of transport is known as ratchet effect.

Usually, ratcheted system in over-damped regime can be recognized as two types depending on whether the potential or force is fluctuating, which are called {\it flashing ratchet} and {\it rocking ratchet} respectively. In most work on classical flashing ratchets, the asymmetric potential is turned on
and off either periodically \cite{ReimannBrownian,cao2004feedback,kinderlehrer2002diffusion,jarillo2018reliability} or at a discrete set of positions \cite{lipowsky2000molecular,kanada2018diffusion}, such that the particle alternates between two states subjected to deterministic control protocols. However, things become much more complicated when we take random control actions into consideration. Such process, of particular interest in this work, was referred to as randomly flashing ratchet, where the Brownian particle randomly jumps between several different processes.

Over the years, investigations on randomly flashing ratchet, though not much abundant, paid predominant attentions to the current or the mean velocity of a Brownian particle in the stationary state \cite{julicher1997modeling, AstumianFluctuation,  rozenbaum2004mechanism, ProstAsymmetric}. Yet it does not suffice to quantify the transport. Predictions of dispersion, or effective diffusion constant, which reveals information on the growth rate of the variance of position, remain inadequate still.

Intuitively, the Brownian particle, in a randomly flashing ratchet, will keep on moving and output mechanical work whenever a weak load is opposed. How effectively the system works should be answered then. Despite extensive studies on the efficiency of diversified ratchet \cite{parrondo2002energetics,suzuki2003rectification,Yu2004Flashing,seifert2012stochastic}, there are few explorations focusing on the thermodynamics of a randomly flashing ratchet and its interactions with the environment.

Our work aims to close these gaps, establishing a general framework of randomly flashing ratchets, which makes it accessible to the derivations of quantities measuring motility and energetics. A Fokker-Planck description of the system will be introduced in \autoref{sec2}, and then we thoroughly discuss the behavior of both dynamic and thermodynamic properties in the stationary state, see \autoref{sec3} and \autoref{sec4} respectively. Two examples, the motility of motor proteins as well as a flashing ratchet with two sawtooth potentials alternating at random, are taken into consideration in \autoref{sec5}. Our results are finally summarized in \autoref{sec6}.

\section{General theoretical framework}\label{sec2}
Let us start with a Brownian particle moving along a one-dimensional track in the over-damped regime and it may stay in $N$ different states. The time evolution of the position $x(t)$, whenever in state $i$, or say potential $U_i(x)$ is switched on, is governed by Langevin equation $\gamma \dot{x}(t)= -U_i^{\prime}(x) - F_i + \sqrt{2\gamma k_B T} \,\xi_i(t)$. Here, $F_i$ is the constant external force (load), $\xi_i(t)$ is the standard Gaussian white noise with $\langle\xi_i(t)\rangle=0$ and $\langle\xi_i(t)\xi_j(t')\rangle=\delta_{i,j}\delta_{t,t'}$, $\gamma$ is the coefficient of viscous drag, $k_B$ is the Boltzmann's constant and $T$ is the absolute temperature. We assume all these $N$ potentials employ the same spatial period $L$, that is $U_i(x) = U_i(x+L)$, and they are different from each other.

Let $\rho_i(x,t)$ be the probability density of finding the particle at position $x$ and in state $i$. Suppose the potential acting on the particle can be switched to another at random, then the motion of particle is captured by the following Fokker-Planck equation on $\rho_i(x,t)$,
\begin{equation}\label{FP}
\begin{aligned}
\parfrac{\rho_i(x,t)}{t} = & \parfrac{}{x}[-a_i(x)\rho_i(x,t) + D\parfrac{}{x}\rho_i(x,t)]\cr
& + \sum_{j = 1}^{N}[w_{ji} \rho_j(x,t) - w_{ij} \rho_i(x,t)],
\end{aligned}
\end{equation}
where $-\infty < x < +\infty$ and $a_i(x) = -\gamma^{-1}[U_i^{\prime}(x) + F_i]$. Here, we assume that transition rate $w_{ij}$ from state $i$ to state $j\ne i$ is nonzero and time homogeneous.

We focus on the stationary state, when the reduced probability density $P_i(x,t)$ becomes independent of time $t$ and the reduced density of average position $S_i(x,t)$ reaches linear growth with time $t$, similar to \cite{derrida1983velocity,zhang2009derivation}. As time $t\to \infty$,
\begin{equation}\label{asz}
\begin{aligned}
P_i(x,t) &= \sum_{k = -\infty}^{+\infty}\rho_i(x+kL,t) \to p_i(x), \\
S_i(x,t) &= \sum_{k = -\infty}^{+\infty}(x+kL)\rho_i(x+kL,t) \\
&\to  v_i(x)t + c_i(x) =: s_i(x,t),
\end{aligned}
\end{equation}
where both $p_i(x)$ and $s_i(x,t)$ satisfy the spatial periodic boundary condition
\begin{equation}\label{peri}
\begin{aligned}
&p_i(0) = p_i(L),&\quad& \difffrac{p_i}{x}(0) = \difffrac{p_i}{x}(L),& \\
&s_i(0,t) = s_i(L, t),&\quad& \parfrac{s_i}{x}(0,t) = \parfrac{s_i}{x}(L,t).&
\end{aligned}
\end{equation}
Notably, the reduced probability density in steady state can be normalized as $\sum_{i = 1}^N \int_0^L  p_i(x) \dx = 1$ due to the normalization of $\rho_i(x,t)$. Similarly, mean position of the particle can be written as $\langle S(t) \rangle =  \sum_{i=1}^N \int_0^L S_i(x,t) \dx$ as an immediate consequence of the definition of $S_i(x,t)$.

In the stationary state, \Eref{FP} can be further recast into differential equations concerning $p_i(x)$ and $s_i(x,t)$ with the ansatz \Eref{asz},{\small
\begin{equation}\label{eq1}
\begin{aligned}
\sum_{j = 1}^{N}[w_{ji}p_j(x) - w_{ij} p_i(x)]	&=  \difffrac{}{x}\left [a_i(x)p_i(x) - D\difffrac{}{x}p_i(x)\right ]\\
&=:\difffrac{J_i(x)}{x},
\end{aligned}
\end{equation}
\begin{equation}\label{eq2}
\begin{aligned}
&\sum_{j = 1}^{N}[w_{ji}s_j(x, t) - w_{ij}s_i(x,t)]
	=\parfrac{}{x}\left [a_i(x) s_i(x,t) \right]+v_i(x)\\
&+2D \difffrac{}{x}p_i(x)- a_i(x)p_i(x)-D\frac{\partial^2}{\partial x^2}s_i(x,t).
\end{aligned}
\end{equation}}

\begin{figure}
	\includegraphics[width=1\linewidth]{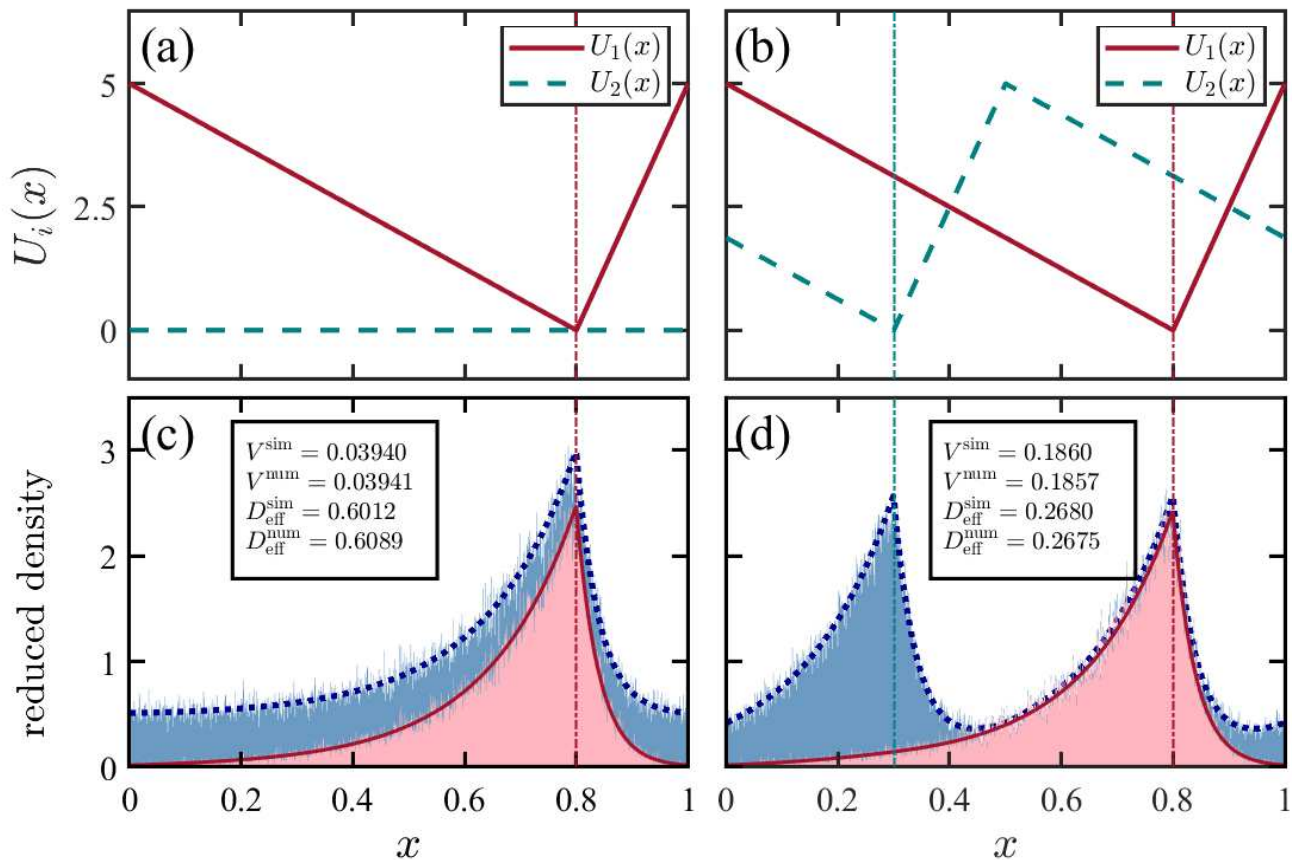}
	\caption{External potentials and the corresponding probability densities over a period. Two different cases are presented. (a)(c): $U_1(x)$ is a sawtooth potential with $\alpha=0.8$, $U_{\max} = 5$, while $U_2(x)\equiv0$ corresponds to the pure diffusion state. (b)(d): Two sawtooth potentials are identical up to a phase difference $\Delta \beta = 0.5$. Asymmetry parameter  $ \alpha_1 = \alpha_2 = 0.8$, potential depth $U^{(1)}_{\max}=U^{(2)}_{\max}=5$. (c) and (d) present the numerical (lines) and simulated results (normalized frequency histogram) of reduced probability densities. Solid line: $p_1(x)$.  Dotted line: $p(x)=p_1(x)+p_2(x)$. The boxes in these two plots display numerical and simulated results for mean velocity and effective diffusion constant. Other parameters: $w_{12}= w_{21} = 1$, $F=0$,  $L=1$, $\gamma = 1$, $k_BT= 1$ and $D=k_BT/\gamma=1$.}
	\label{Simu}
\end{figure}

Now, we are prepared to obtain $p_i(x)$ through solving the second-order homogeneous differential equations \Eref{eq1} with periodic boundary conditions and the normalization condition. Simulations of a single Brownian particle in randomly flashing potential validate the theoretical predictions suggested by Fokker-Planck equations, see \Fref{Simu}. In \Fref{Simu}(a) and (c), $U_1(x)$ is a piecewise linear potential while $U_2(x) \equiv 0$ implies a free diffusion. In \Fref{Simu}(b) and (d),  $U_1(x)$ and $U_2(x)$ are identical sawtooth potentials up to a phase difference $\Delta \beta= \beta_2-\beta_1$, {\it i.e.}, $U_2(x)=U_1(x-\Delta \beta L)$. Here, the sawtooth potential reads
\begin{equation}\label{potential}
U(x) =
\begin{cases}
U_{\max}[\alpha L-(x-\beta L)] / (\alpha L), \\
\qquad\textrm{for}\quad 0\leq\!\!\!\!\mod (x-\beta L,L) <\alpha L,\\
U_{\max}[(x-\beta L) -\alpha L] / (L-\alpha L),\\
\qquad\textrm{for}\quad \alpha L\leq\!\!\!\!\mod (x-\beta L,L) < L,
\end{cases}
\end{equation}
where $U_{\max}$ is the potential depth, $\alpha$ is the asymmetry parameter, $L$ describes the period of potential and $\beta$ is the phase shift.

\section{Velocity and dispersion}\label{sec3}
It is straightforward to obtain the mean velocity in the stationary-state limit starting with the definition. See Sec.~S1 in {\bf Supplemental Materials} for details.
\begin{equation}\label{eqv}
V := \lim_{t \to +\infty}\difffrac{\langle S(t) \rangle}{t} = \sum_{i=1}^N \int_0^L a_i(x)p_i(x) \dx,
\end{equation}
which coincides with the classical result $V = JL$ as the total current $J=\sum_{i=1}^{N}J_i(x)$ is $x$-independent. The effective diffusion constant is defined by \cite{derrida1983velocity,zhang2009derivation},
\begin{equation*}
D_{\rm eff} := \lim_{t \to +\infty} \frac{1}{2}\difffrac{\left[\langle S^2(t)  \rangle- \langle S(t) \rangle^2\right]}{t}.
\end{equation*}
After a short calculation, we obtain
\begin{equation*}
D_{\rm eff} = D + \sum_{i=1}^N \int_0^L a_i(x)s_i(x,t) \dx - V \sum_{i=1}^N\int_0^L s_i(x,t)\dx .
\end{equation*}
It can be shown that terms proportional to time $t$ in $D_{\rm eff}$ can be canceled out. Besides, $c_i(x)$ in $s_i(x,t)$ actually depends on an undetermined constant which will also cancel out in the final expression for $D_{\rm eff}$. With these two cancellations, we arrive at
\begin{equation}\label{eqD}
D_{\rm eff} = D + \sum_{i=1}^N \int_0^L [a_i(x)-V]c_i(x) \dx.
\end{equation}
See Sec.~S2 in {\bf Supplemental Materials} for detailed derivation.
\Fref{Simu}(a) and (c) show that the theoretical predictions of mean velocity $V$ and effective diffusion constant $D_{\rm eff}$, \Eref{eqv} and \Eref{eqD}, agree with the simulations well, and the Brownian particle is more likely to be found around the valley $\alpha L$ of potential $U_1(x)$. We take $N=2$ in \Fref{Simu} and all other figures below.
\begin{figure}[htbp]
	\includegraphics[width=1\linewidth]{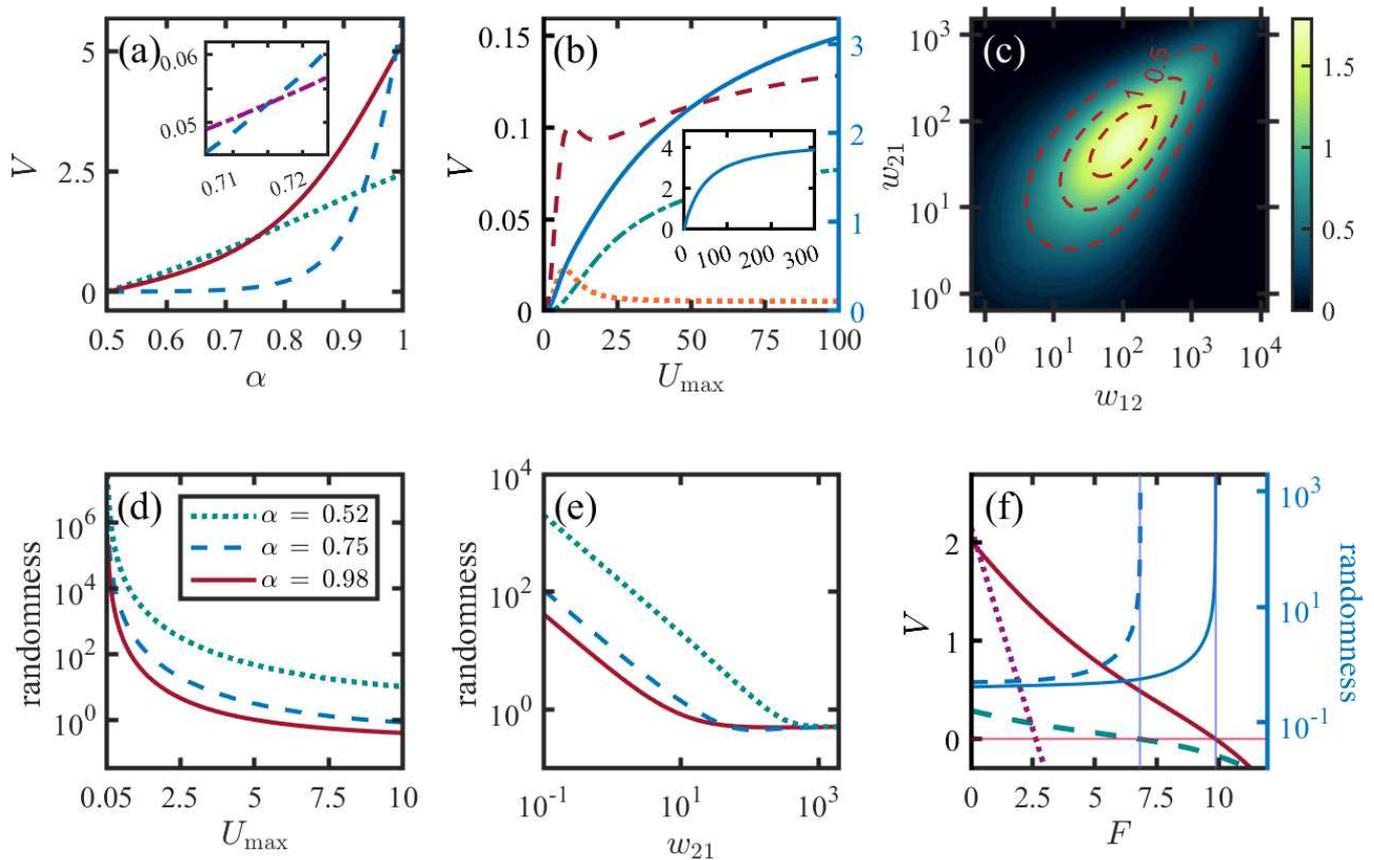}
	\caption{Dynamic properties of a randomly flashing ratchet. (a) Dependence of mean velocity on asymmetric parameter $\alpha$, where $U_{\max} = 20$, $w_{12}=100$. Dotted line: $w_{21}=25$. Solid line: $w_{21}=100$. Dashed line: $w_{21}=500$. Inset: Dashed line is exact the one in (a). Dash-dotted line: $U_{\max} = 5$, $w_{12}=100$, $w_{21}=500$. (b) Dependence of mean velocity on potential depth $U_{\max}$. $w_{12}=100$. Dash-dotted line: $\alpha = 0.6$, $w_{21}=1$. Dotted line: $\alpha = 0.6$, $w_{21}=500$. Dashed line: $\alpha = 0.75$, $w_{21}=500$. Solid line(right axis): $\alpha = 0.9$, $w_{21}=500$. Inset: the solid line in (b) plotted over a wider range of $U_{\max}$. (c) Contour map of mean velocity as a function of the two transition rates $w_{12}$ and $w_{21}$, with $U_{\max} = 20$, $\alpha = 0.8$. (d) Randomness as a function of potential depth $U_{\max}$, where $w_{12} = w_{21}=100$. (e) Dependence of randomness on transition rates, where $U_{\max} = 20$ and $\alpha=0.8$. Solid line: $w_{12}=10^0$. Dashed line: $w_{12}=10^2$. Dotted line: $w_{12}=10^3$. (f) Dependence of mean velocity (left axis) and randomness (right axis) on load $F$. Dashed line: $w_{21}=300$, $\alpha=0.75$. Solid line: $w_{21}=300$, $\alpha=0.9$. Dotted line: $w_{21} = 30$, $\alpha = 0.9$. $U_{\max} = 20$ and $w_{12} = 100$ hold. Other parameters: $L=1$, $\gamma=1$, $k_BT = 1$, $D=k_BT/\gamma=1$ and $F=0$ except (f).}
	\label{Velo}
\end{figure}

To discuss the mobility, we consider the flashing ratchet with random alternations of two potentials, a sawtooth potential $U_1(x)$ and a flat one $U_2(x)\equiv0$, provided no load. This is depicted in \Fref{Simu}(a). As expected, the mean velocity rises with $\alpha$ approaching $1$, due to the increase of possibility to slide down the slope of sawtooth potential $U_1$. See \Fref{Velo}(a). Besides, the mean velocity becomes more sensitive to  asymmetry parameter  $\alpha$ with more frequent transition $2 \to 1$, since the particle will have more probability to move in the sawtooth potential.

Higher mean velocity will also be realized provided with a deeper sawtooth potential, and the mean velocity displays asymptotic behavior when potential depth $U_{\max}$ is extremely large, see \Fref{Velo}(b). This implies that increasing input energy may not be always effective to improve the performance of the randomly flashing ratchet as a mechanical machine. Meanwhile, the dependence of mean velocity on potential depth $U_{\max}$ may become nonmonotonic with large transition rates $w_{21}$ and $\alpha$ close to $0.5$. Recall the expression for mean velocity given in \Eref{eqv} and we find it is the sum of sliding velocity $a_i(x)$ weighted by probability $p_i(x)$ accordingly. Large $w_{21}$ implies the particle spends much more time moving in sawtooth potential $U_1(x)$. With the increase of potential depth $U_{\max}$, the increment of velocity arising from sliding down potential towards $+x$ direction may not compensate for that towards $-x$ direction for $\alpha$ close to $0.5$, and the mean velocity may therefore slow down as the sawtooth potential gets deeper. See the dotted line and the dashed line in \Fref{Velo}(b) and the inset in \Fref{Velo}(a).

Apart from the profile of the sawtooth potential $U_1(x)$, transition rates between potentials also make difference in the mean velocity, which is actually the key point to the directional transport.  For the specific example aforementioned, the mean velocity in the steady state vanishes as either $w_{12}$ or $w_{21}$ goes zero (or goes infinity),  meaning the particle being trapped in one of the two potentials --- either case leads to zero displacement on average in the stationary state. Meanwhile, the mean velocity vanishes with extremely frequent switching of potentials, since there is little time for the particle to slide down to the valley in the sawtooth potential or to diffuse in the  flat potential. Thus, a peak can be observed in \Fref{Velo}(c), which presents the dependence of mean velocity on transition rates. It is the moderate flipping rates that bring high mean velocity. Intuitively, the particle gains large velocity only when the system can effectively \lq\lq feel" the two potentials. Such relations between mean velocity and transition rates can also be found in \Fref{Velo}(a), if we take $\alpha \gtrsim 0.8$ fixed.

In addition to mean velocity, the dispersion, quantified by effective diffusion constant $D_{\rm eff}$, plays an important role in describing the performance of a randomly flashing ratchet. Unfortunately, $D_{\rm eff}$ presents really complicated behavior when we discuss its dependence on flipping rates or parameters involving the sawtooth potential, making it inaccessible to intuitive insight. Thus, we focus on the dimensionless function randomness instead, defined as
{\small
\begin{equation*}
R = \frac{D_{\rm eff}}{VL},
\end{equation*}}
to detect the degree of randomness (or determinacy) of the motion of a Brownian particle in the randomly switching potentials. \Fref{Velo}(d) shows that randomness $R$ decreases with both potential depth $U_{\max}$ and asymmetry (measured by $|\alpha-1/2|$). Intuitively, randomness mainly comes from the free diffusion process in flat potential $U_2(x)\equiv0$ while determinacy is generated by the asymmetric sawtooth potential $U_1(x)$. Randomness $R$, therefore, increases with transition rate $w_{12}$ but decreases with $w_{21}$, see \Fref{Velo}(e).

For a Brownian particle under constant external load $F_1=F_2=F$, \Fref{Velo}(f) shows that mean velocity $V$ decreases with load $F$ while randomness $R$ increases, and randomness $R$ goes infinity as the load $F$ tends to {\it stall force} $F_{\rm stall}$, which is defined by $V(F_{\rm stall})=0$ and marked by vertical lines in \Fref{Velo}(f). Strong {\it stall force} $F_{\rm stall}$ can be achieved with large transition rate $w_{21}$ or strong potential asymmetry. The mean velocity $V$ exhibits high sensitivity to load $F$ in the low frequency limit of transition $2\to1$ inferring longer average time $1/w_{21}$ to diffuse freely in potential $U_2(x)\equiv0$. That is when linear dependence arises, see the dotted line in \Fref{Velo}(f).

\section{Power and efficiency}\label{sec4}
\begin{figure}[htbp]
	\includegraphics[width=1\linewidth]{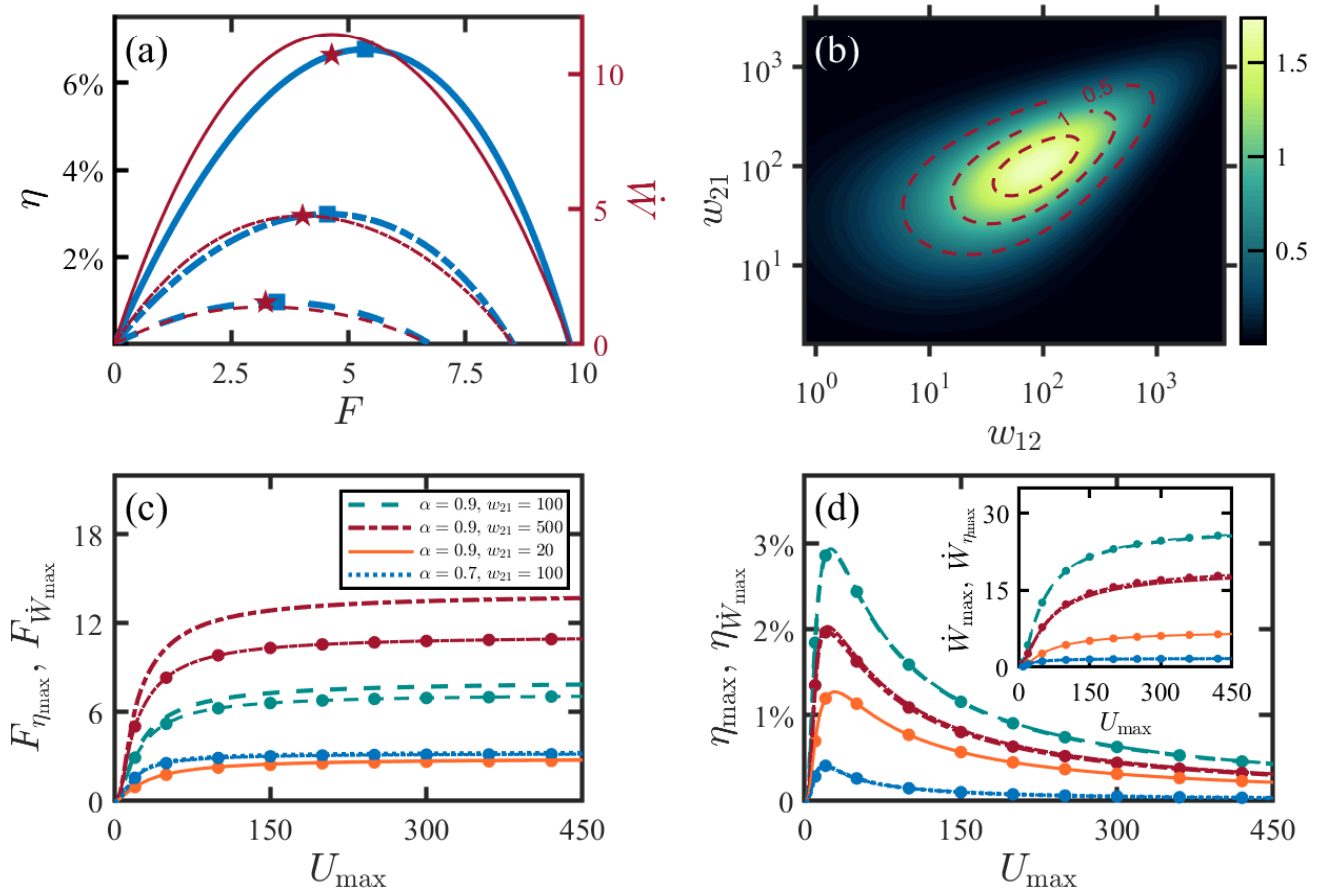}
	\caption{Energetics of a randomly flashing ratchet. (a) Load $F$ dependence of efficiency $\eta$ (thick line, left axis) and output power $\dot{W}$ (thin lines, right axis), with $U_{\max}=20$, $w_{12}=100$ and $w_{21}=200$. Dashed line: $\alpha =0.8$. Dash-dotted line: $\alpha=0.9$. Solid line: $\alpha=0.98$. The maximum of efficiency is marked by \textcolor[RGB]{0,115,189}{$\blacksquare$} and the efficiency at maximum power is marked by \textcolor[RGB]{163,20,46}{$\bigstar$}. (b) Maximum power $\dot{W}_{\max}$ as a function of transition rates $w_{12}$ and $w_{21}$ with $U_{\max}=20$, $\alpha=0.8$. (c) Comparison of external load at maximum efficiency ($F_{\eta_{\max}}$, lines) to that at maximum power ($F_{\dot{W}_{\max}}$, lines with $\bullet$). For all lines $w_{12} =100$ while $\alpha$ and $w_{21}$ are illustrated in the legend. (d) Maximum efficiency ($\eta_{\max}$, lines) and efficiency at maximum power ($\eta_{\dot{W}_{\max}}$, lines with $\bullet$), as the change of potential depth $U_{\max}$. Corresponding output power are presented in the inset. Parameters $\alpha$, $w_{12}$ and $w_{21}$ for each line are the same as those in (c). Other parameters: $k_BT=1$, $\gamma=1$ and $D=k_BT/\gamma=1$.}
	\label{thermo}
\end{figure}

Randomly flashing ratchets can be regarded as stochastic thermodynamic machines. The Brownian particle gains energy everytime the potential acting is flipped to another, and then part of the energy will be converted to mechanical work against load whereas the rest is dissipation due to the nontrival viscocity. In the stationary state, the {\it input energy} per unit time generally reads \cite{ParrondoEfficiency,Yu2004Flashing,parrondo2002energetics}
{\small
\begin{equation}
\begin{aligned}
&\dot{E}_{in}
= \sum_{i,j=1}^N \int_0^L  U_i(x)[w_{ji}p_j(x)-w_{ij}p_i(x)] \dx \\
=& \sum_{1\leq i < j \leq N} \int_0^L [U_j(x)-U_i(x)][w_{ij}p_i(x)-w_{ji}p_j(x)] \dx,
\end{aligned}
\end{equation}}
and the {\it power} $\dot{W}$ is given by
\begin{equation}\label{eqPower}
\dot{W} = \sum_{i=1}^N F_i \int_0^L J_i(x) \dx = \sum_{i=1}^N F_i \int_0^L a_i(x)p_i(x) \dx.
\end{equation}
Particularly, $\dot{W} = FV$ if $F_i \equiv F$ for $1\leqslant i \leqslant N$. Thus, the energy efficiency is $\eta = \dot{W}/\dot{E}_{in}$, indicating to which degree input energy is converted to mechanical work.

Obviously, zero load implies no mechanical work and the power $\dot{W}=0$. Meanwhile, the increasing load $F$ reduces the mean velocity $V$ (\Fref{Velo}(f)) and the power $\dot{W}$ finally vanishes at the stalling force $F_{\rm stall}$.  As a result, the maximum power $\dot{W}_{\max}$ appears when load $F$ varies from 0 to $F_{\rm stall}$, and the same behavior is observed for efficiency $\eta$, see \Fref{thermo}(a). Although a steeper sawtooth potential, which is suggested by greater $U_{\max}$ or stronger asymmetry, guarantees not only a stronger force maximizing power $\dot{W}$, denoted as $F_{\dot{W}_{\max}}$, but also a stronger force maximizing efficiency $\eta$, denoted as $F_{\eta_{\max}}$, there exists saturation amount, see \Fref{thermo}(c). In other words, if we keep the thermodynamic machine working at the maximum efficiency/power, increasing potential depth and asymmetry both contribute to the success in overcoming stronger load, but the enhancement is limited. Though $F_{\eta_{\max}}\ge F_{\dot{W}_{\max}}$, the corresponding efficiencies, {\it i.e.}, maximum efficiency $\eta_{\max}$ and efficiency at maximum power $\eta_{\dot{W}_{\max}}$, are almost the same, see \Fref{thermo}(d). They both increase first and then decrease with potential depth $U_{\max}$,  presenting the maxima that depends on asymmetry parameter and transition rates. Also, there is no significant difference between the power at $F_{\eta_{\max}}$ and $F_{\dot{W}_{\max}}$, denoted by $\dot{W}_{\eta_{\max}}$ and $\dot{W}_{\max}$ respectively, and the dependence on  $\alpha$ and $U_{\max}$ display similar behavior to those of $F_{\eta_{\max}}$ and $F_{\dot{W}_{\max}}$, see the inset in \Fref{thermo}(d).

 Similar to the mean velocity, a peak can be found in the contour map of maximum power $\dot{W}_{\max}$, regarded as a function of transition rates, which is illustrated in \Fref{thermo}(b) and confirmed by the inset in \Fref{thermo}(d).

\section{Case Study}\label{sec5}
So far, we have introduced a general framework which allows us to study the dynamic and thermodynamic properties of a randomly flashing ratchet. We now discuss two examples: applications to molecular motors and a 2-state system with two sawtooth potentials.

\subsection{Motility of molecular motors}
Biological molecular motors, usually known as motor proteins, move along cytoskeleton filaments unidirectionally \cite{Howard2001,Vale2003}. The randomly flashing ratchet model will be applied to the characterization of the mechanochemical process of kinesin and dynein, serving as an example. We consider the simple case where kinesin/dynein always walks along microtubules (MT) without interruption of detachment and attachment. Roughly speaking, kinesin/dynein are mainly in two states. (i)Free diffusion state, when one head binds to the MT yet the other, binding ADP, detached from MT and freely diffuses around. (ii)Power stroke state with two head binding to MT simultaneously. With ATP binding to the front head, neck linkers of motor will swing forward and stick closely to the MT, which then leads to a forward motion of the motor. The mechanochemical process of a motor is depicted by the randomly flashing ratchet, where a Brownian particle in flat potential $U_2(x)\equiv0$ accounts for state (i), state (ii) reveals the motion in sawtooth potential $U_1(x)$, and the transitions between two potentials are triggered by chemical reactions, ATP binding and hydrolysis (the release of phosphate ion) \cite{Zhang2008,Kolomeisky2015}.

\begin{figure}[htbp]
	\includegraphics[width=1\linewidth]{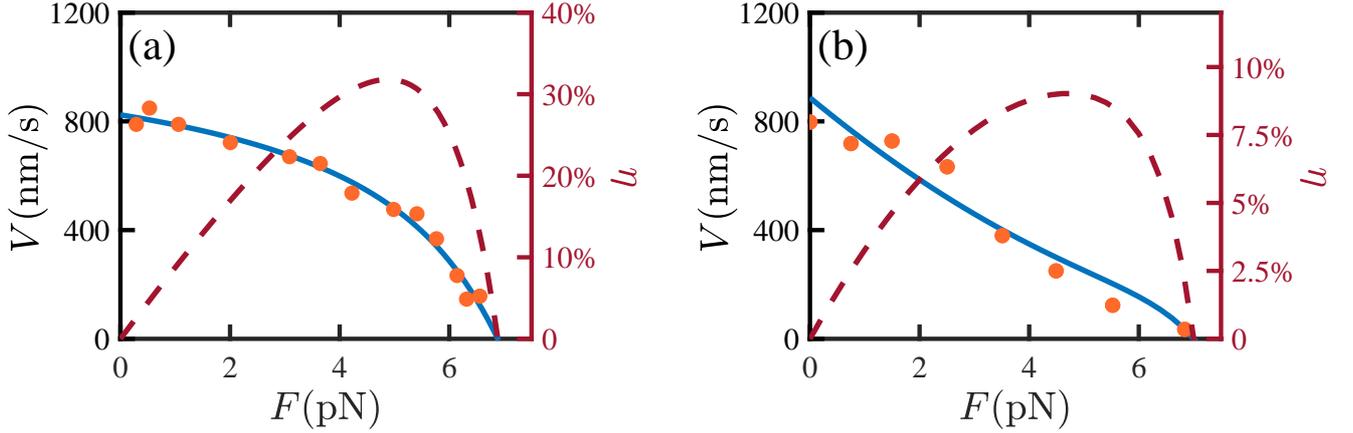}
	\caption{Force dependent velocity $V$ (solid line, left axis) and efficiency $\eta$ (dashed line, right axis) of kinesin (a) and dynein (b). (a) Experimental data are measured at 2 mM ATP for conventional kinesin purified from squid optic lobe \cite{VisscherSingle}. Fitted parameters: $U_{\max} = 17.45 \ k_BT$, $\alpha=0.9573$, $D=8.259\times10^4 \ \text{nm}^2/\text{s}$, $w_{12} = 3.934\times 10^{3}\text{ s}^{-1}$, $w_{21}= 3.285\times 10^2\text{ s}^{-1}$. (b) Experimental data are measured at 1 mM ATP for cytoplasmic dynein purified from porcine brain \cite{toba2006overlapping}. Fitted parameters: $U_{\max} = 22.95 \ k_BT$, $\alpha =0.9823$, $D= 1.108 \times 10^{5} \text{ nm}^2/\text{s}$, $w_{12}=1.212\times10^3 \text{ s}^{-1}$, $w_{21}= 8.361\times10^{5}\text{ s}^{-1}$. Other parameters: $L=8.2$ nm (step size ), $k_BT=4.12\ \text{nm}\cdot\text{pN}$.}
	\label{motor}
\end{figure}

\Fref{motor} exhibits the force dependent velocity $V$ and efficiency $\eta$ of conventional kinesin and cytoplasmic dynein with several parameters fitted. As expected, for both kinesin and dynein, the mean velocity $V$ decreases with external load $F$ while the efficiency $\eta$ increases first and then decreases to zero as the load $F$ increases from 0 to the stall force $F_{\rm stall}$. We should notice that kinesin always moves towards the plus end of MT, and the directionality  is so strong that kinesin can hardly be pulled back, even if the load exceeds the stall force $F_{\rm stall}$. Accordingly, we take the external force $F_2\equiv0$ in the diffusion state of kinesin. However, experiments reveal that the motion of dynein is not so robust. Although dynein moves to the minus end of MT on average, it may walk backwards occasionally and stronger load, below the stall force $F_{\rm stall}$, suggests higher backward frequency. External load is exerted to the motor in each state equally, {\it i.e.}, $F_1\equiv F_2\equiv F$, and thus kinesin works more effectively than dynein. See \Fref{motor}.

\subsection{Randomly flashing ratchet with two sawtooth potentials }
\begin{figure}[htbp]
	\includegraphics[width=1\linewidth]{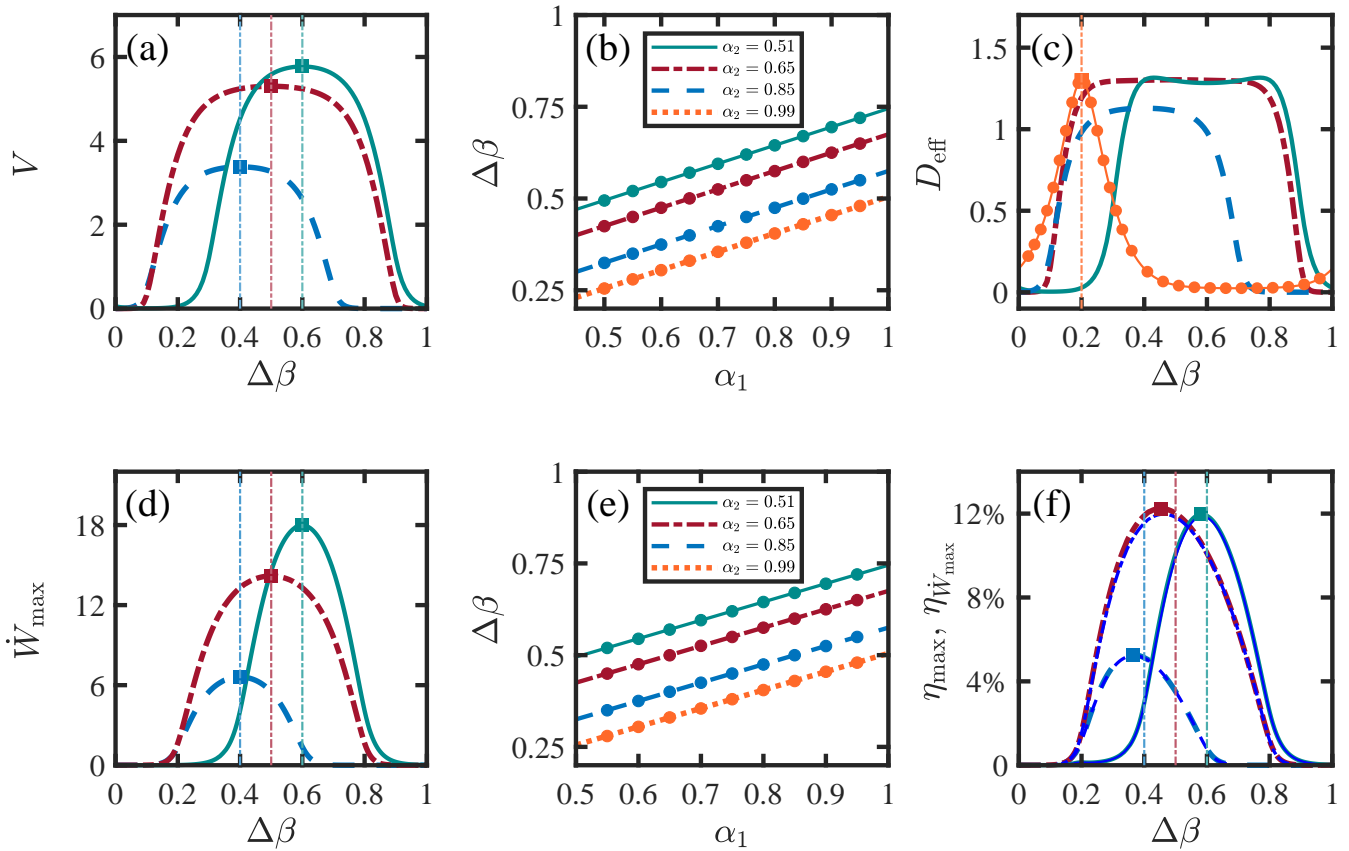}
	\caption{Properties of the randomly flashing ratchet with two sawtooth potentials.  (a) Mean velocity $V$ as a function of phase difference $\Delta \beta$, with the maximum of $V$ marked by $\blacksquare$. Solid line: $\alpha_1= 0.9$,  $\alpha_2=0.7$. Dash-dotted line: $\alpha_1= \alpha_2 = 0.9$.  Dashed line: $\alpha_1= 0.7$,  $\alpha_2=0.9$.  (b) The asymmetry parameter $\alpha_1$ dependent optimal value of $\Delta \beta$, when $V$ reaches its maximum as marked by $\blacksquare$ in (a). Data $(\alpha_1, \Delta \beta)$ marked by $\bullet$ are obtained by maximizing $V$ (see the points marked by $\blacksquare$ in (a)), while lines are plotted according to relation $\Delta \beta = (1-\alpha_2+\alpha_1)/2$. (c) Dependence of $D_{\rm eff}$ on $\Delta \beta$. Solid line with $\bullet$: $\alpha_1=0.2$, $\alpha_2=0.8$. Other lines corresponds to (a). (d) Dependence of maximum power $\dot{W}_{\max}$ on $\Delta \beta$. Solid line: $\alpha_1=0.8$, $\alpha_2=0.6$. Dash-dotted line: $\alpha_1=\alpha_2=0.8$. Dashed line: $\alpha_1=0.6$, $\alpha_2=0.8$. (e) Parameter $\alpha_1$ dependent optimal value of $\Delta \beta$ which maximizes $\dot{W}_{\max}$. Similar as in (b), each line corresponds to $\Delta \beta = (1-\alpha_2+\alpha_1)/2$ while each $\bullet$ marks $(\alpha_1, \Delta \beta)$ that maximizes $\dot{W}_{\max}$. (f) Dependence of $\eta_{\max}$ (thick lines) and $\eta_{\dot{W}_{\max}}$ (thin lines) on $\Delta \beta$. In (a, c, d, f), each $\blacksquare$ marks the maxima while the vertical line corresponds to $\Delta \beta = (1-\alpha_2+\alpha_1)/2$. Other parameters: $U_{\max}^{(1)}=20$, $U_{\max}^{(2)}= 10$, $w_{12}=20$, $w_{21} = 50$, $L=1$, $\gamma=1$, $k_BT=1$ and $D= k_BT/\gamma=1$.}
	\label{2saw}
\end{figure}

We now discuss the two-state randomly flashing ratchet where $U_1$ and $U_2$ are both sawtooth potentials of period $L$. $U_i \ (i=1,2)$ employs potential depth $U_{\max}^{(i)}$, asymmetry parameter $\alpha_i$ and phase shift $\beta_i$. See \Eref{potential} for explicit expressions. Both numerical calculations and simulations show that either of the two valleys indicates more probability of finding the Brownian particle in the stationary state, see \Fref{Simu}(b) and (d). Without loss of generality, we define the phase difference as $\Delta \beta=\mod(\beta_2-\beta_1, 1)$ with initial phase $\beta_1=0$, and the discussions in the following focus on the effects of phase difference $\Delta \beta$.

We consider the system with $0.5<\alpha_1, \alpha_2<1$ except as noted, promising a positive mean velocity, and we then find there is something amazing at $\Delta \beta = [1- (\alpha_2-\alpha_1)] / 2$ --- it is exactly the point maximizing both mean velocity $V$ and maximum power $\dot{W}_{\max}$, even with two sawtooth potentials different in potential depth and asymmetry and unequal transition rates. See \Fref{2saw}(a, b, d, e). In terms of the effects on $D_{\rm eff}$, $\Delta \beta = [1- (\alpha_2-\alpha_1)] / 2$ is still a most special point, although there may exist more than one peaks as $\Delta \beta$ varies in $[0,1]$, see \Fref{2saw}(c). For specific cases with $\alpha_1 + \alpha_2 =1$, only $\Delta \beta = \alpha_1$ maximizes $D_{\rm eff}$, and it creates a situation where the peaks of one potential match the valleys of the other, see the solid line with $\bullet$ in \Fref{2saw}(c). However, the phase difference $\Delta \beta$ maximizing $\eta_{\max}$ is slightly less than that maximizing $\dot{W}_{\max}$, though the $\Delta \beta$-dependent maximum efficiency $\eta_{\max}$ approximates the efficiency at maximum power $\eta_{\dot{W}_{\max}}$ quite well (\Fref{2saw}(f)). When $\Delta \beta=0$ (or equivalently $\Delta \beta=1$), the Brownian particle seemingly alternates between the minima of the two potentials in the same period and can hardly cross the potential barriers nearby, resulting in negligible mean velocity $V$ and effective diffusion constant $D_{\rm eff}$, see \Fref{2saw}(a, c). So do the maximal output power and energetic efficiency, see \Fref{2saw}(d, f).

The dependence of mean velocity $V$ on transition rates $w_{12}$ and  $w_{21}$ behaves similarly as that of the usual two-state model discussed in previous sections, \ref{sec3} and \ref{sec4}, while the presence of double peaks can be observed in the plots of $D_{\rm eff}$, see Fig.~S1 in  {\bf Supplemental Materials}.

\section{Conclusions and Remarks}\label{sec6}
Randomly flashing ratchet is characterized through coupled Fokker-Planck equations. Expressions of quantities measuring dynamics and thermodynamics are determined, and the simulations validate these theoretical predictions then. General performance of a randomly flashing ratchet is thoroughly studied with respect to system parameters. Motility of motor proteins, kinesin and dynein, is further elucidated with the solvable model proposed in this work. The randomly flashing ratchet with two sawtooth potentials alternating is discussed finally, and the optimal performance of the ratchet, serving as a thermodynamic machine, arises when $\Delta \beta = [1-(\alpha_2-\alpha_1)]/2$.

\end{document}